\pdfoutput=1
\documentclass[conference,compsoc]{IEEEtran}

\usepackage{amsmath,amssymb}
\usepackage{graphicx}
\usepackage{booktabs}
\usepackage{enumitem}
\usepackage[nocompress]{cite}
\usepackage{url}
\usepackage{xurl}
\usepackage[hidelinks]{hyperref}
\usepackage{microtype}
\usepackage{xcolor}
\usepackage{placeins} %

\usepackage{iftex}

\hypersetup{
    pdfauthor={Liangyi Huang, Zichen Liu, Fei Shao, Xusheng Xiao},
    pdftitle={LODESTONE: Linking System Events to Threat Intelligence for Evidence-Grounded Cyber Attack Investigation}
}

\title{LODESTONE: Linking System Events to Threat Intelligence for Evidence-Grounded Cyber Attack Investigation}

\author{
\IEEEauthorblockN{Liangyi Huang}
\IEEEauthorblockA{Arizona State University\\
lhuan139@asu.edu}
\and
\IEEEauthorblockN{Zichen Liu}
\IEEEauthorblockA{Arizona State University\\
zliu396@asu.edu}
\and
\IEEEauthorblockN{Fei Shao}
\IEEEauthorblockA{Case Western Reserve University\\
fxs128@case.edu}
\and
\IEEEauthorblockN{Xusheng Xiao}
\IEEEauthorblockA{Arizona State University\\
xusheng.xiao@asu.edu}
}

\begin{document}
\maketitle

\begin{figure*}[t]
    \centering
    \includegraphics[width=\textwidth]{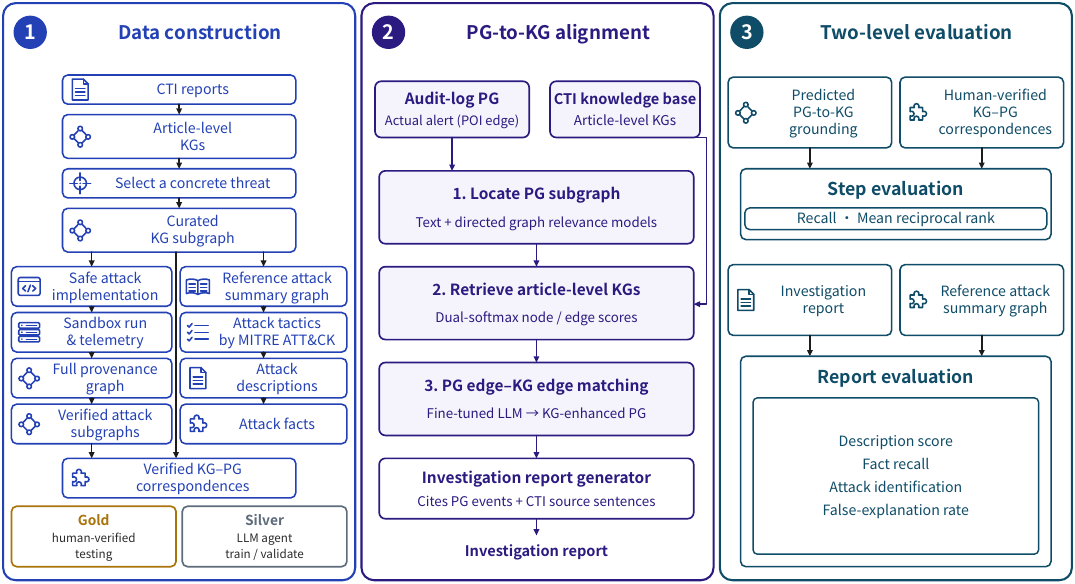}
    \caption{LODESTONE's data construction, investigation pipeline, and evaluation.}
    \label{fig:overview}
\end{figure*}

\begin{abstract}
System provenance graphs help reconstruct attacks through causal dependencies, but these connections alone do not explain how events accomplish an adversary’s goal. Cyber threat intelligence (CTI) provides the necessary security knowledge, yet linking it to execution evidence requires bridging differences in meaning and granularity between events and attack relations amid scarce labels and sparse relevant evidence. We present LODESTONE, a cross-graph alignment system that links observed system events to CTI attack relations to support evidence-grounded attack reports.

LODESTONE uses a novel CTI-driven training-data engine and an expert-verified benchmark of edge-level correspondences and reference attack reports. The engine pairs real CTI knowledge graphs with provenance graphs from isolated executions. A large language model (LLM) agent recursively improves correspondences by checking execution evidence and repairing retrieval, parsing, and reproduction failures; experts independently verify benchmark correspondences. LODESTONE combines text cross-encoder and directed-graph relevance scores to extract compact alert-centered subgraphs, retrieves article-level knowledge graphs through node- and edge-level dual-softmax assignment, and uses a fine-tuned 4B-parameter LLM to align event combinations with CTI attack steps. These many-to-many links support attack reports citing observed events and CTI source sentences.

We evaluate LODESTONE on 27 expert-verified attack cases amid continuous benign activity from real applications, using 713 CTI articles. LLM judgments aligned with multiple human experts yield an attack-step description score of 82\%, and attack identification reaches 91\%, versus 73\% and 32\% using the same generator with provenance alone. Baseline pipelines and coding agents achieve description scores of at most 57\%. Human-verified correspondences further demonstrate advantages in balancing attack-edge coverage with subgraph size, CTI retrieval, and cross-graph alignment.

\end{abstract}

\section{Introduction}
\label{sec:introduction}

Advanced cyber attacks, including advanced persistent threat (APT) attacks, combine multiple techniques into multistep executions whose malicious intent can be obscured by benign-looking activities~\cite{milajerdi2019holmes}. Detecting an isolated suspicious event therefore does not reconstruct the attack or explain how its steps are connected. Provenance analysis (PA) addresses this problem by constructing a provenance graph (PG) from system-call audit logs to capture dependencies among system activities~\cite{king2003backtracker,hossain2017sleuth}. The PG supplies the execution context needed to investigate an alert. A point-of-interest (POI) edge represents the actual system event that triggered an upstream alert and anchors the investigation. Backward traversal identifies preceding events and possible entry points, while forward traversal reveals subsequent activity and potential consequences~\cite{king2003backtracker,liu2018priotracker}. Previous studies further identify attack sequences or reduce unrelated provenance to focus the investigation~\cite{alsaheel2021atlas,fang2022depimpact,xu2022depcomm}. This context connects suspicious events across processes and resources, but dependencies alone do not establish which event combinations implement an attack step or how those steps accomplish an adversary's goal. Investigators also need security knowledge that explains the observed behavior.

Cyber threat intelligence (CTI) provides an external source of that security knowledge. Indicators of compromise identify artifacts such as malicious names, addresses, and domains~\cite{liao2016iocgame}; CTI articles additionally describe behaviors, relationships, and tactics observed in real incidents. These descriptions can be extracted into article-level knowledge graphs (KGs)~\cite{satvat2021extractor,li2022attackg,huang2024ctikg}. Previous studies establish the value of connecting such knowledge to provenance through inexact matching with CTI query graphs~\cite{milajerdi2019poirot} and learned graph representations for matching known attack behavior~\cite{wei2021deephunter}. However, applying these approaches to article-derived knowledge exposes a gap: their matching assumes a shared vocabulary of typed system entities and events, whereas the article-level KGs used here keep high-level entities, such as malware families, and free-form attack relations. This mismatch limits retrieval by learned graph representations; in our evaluation, the inexact matcher finds partial alignments, but none reaches its published acceptance threshold. Moreover, to our knowledge, no dataset provides expert-verified correspondences between individual article relations and the system events that carry them out, so such links can be neither learned nor measured.

To overcome this fundamental limitation in connecting CTI knowledge to provenance evidence for cyber threat investigation, we propose \textbf{PG--KG graph alignment}: \textit{grounding expert knowledge of attack behavior in observed execution evidence}. Given an alert-centered PG and a collection of article-level KGs, the task is to identify which combinations of low-level system events support higher-level CTI attack steps despite differences in vocabulary and granularity. This requires determining which articles explain the observed activity: the relevant articles are initially unknown, several may jointly explain the execution, and the knowledge base may contain no adequate explanation. The alignment therefore establishes explicit event-to-relation links while leaving unsupported behavior unmatched. Each link retains the original PG event identifiers and the CTI source sentences associated with its KG relation, allowing investigators to trace an interpretation to both its execution evidence and its documented security meaning. We call this \emph{PG-to-CTI grounding}. Building on these links, an \emph{attack summary graph} (ASG) organizes the supported attack steps and their progression into a case-level report that cites both observed events and CTI source sentences, helping investigators explain the activity and reconstruct the attack. Achieving this evidence-grounded investigation through PG--KG graph alignment requires addressing four key research challenges.

\noindent\emph{C1: Labeled attack traces are scarce amid abundant benign activity.}
Routine operation produces extensive audit data, but realistic attacks with fine-grained ground truth are difficult to obtain. Previous work learns from benign executions and reports that attacks occupy less than $0.001\%$ of the audit data in its DARPA evaluation~\cite{han2020unicorn}; recent dataset research also identifies outdated attacks, incomplete chains, and coarse labels as persistent limitations~\cite{cai2026building}. Grounding requires more than an attack label: the same case must provide CTI source passages, a reproducible execution, and edge-level links between the KG and PG. Benign traces cannot supply these correspondences, and unconstrained execution of reported malicious behavior is unsuitable for collecting them.

\noindent\emph{C2: System dependencies do not supply security expert knowledge.}
A file read or network connection may belong to routine administration or to credential theft and exfiltration. Even after provenance reduction isolates a small subgraph~\cite{fang2022depimpact,xu2022depcomm}, recognizing the attack goal requires knowledge of the technique and its surrounding steps. CTI articles encode this expertise, but retrieving a semantically similar sentence does not establish that the observed combination of events implements it. An explanation must identify concrete supporting evidence, distinguish unsupported interpretations, and preserve its connection to the report that supplied the security meaning.

\noindent\emph{C3: PG--KG matching crosses semantic and granularity gaps.}
PG edges describe low-level process behavior~\cite{king2003backtracker}; KG edges express higher-level attack steps~\cite{huang2024ctikg}. Credential theft may involve opening a credential file, reading secrets, and staging output. None of these individual event descriptions need resemble the CTI wording, while a generic read operation can resemble many unrelated relations. A KG step may require several PG edges, and a PG event may support several relations from different articles. Some relevant PG events have no CTI counterpart, and some KG relations leave no observable system event. Matching must therefore be many-to-many and partial; one-to-one semantic assignment is insufficient.

\noindent\emph{C4: Relevant evidence is sparse in both graphs and reports.}
Attack events are a small subset of a PG dominated by benign activity. Causal reachability from an alert can expand through common processes and files without identifying attack relevance. The knowledge base creates a second search problem: hundreds of articles may share generic techniques, but only a few explain the current execution.
A system must recover compact evidence, select relevant KGs, and acknowledge insufficient support when no retrieved article explains the alert.

\noindent\textbf{LODESTONE.}
We present LODESTONE, a graph alignment system that takes a PG, its POI edges, and a collection of article-level KGs as input and produces event-to-relation links and an ASG for attack investigation (Figure~\ref{fig:overview}). A CTI-driven data engine supplies the labeled cases needed to learn and evaluate the alignment pipeline.

Starting from CTI articles, the data engine composes compatible techniques into multistage scripts, records their system effects in isolated execution environments, and pairs the resulting PGs with source-grounded KG--PG correspondences (Challenge \emph{C1}). Human experts verify the test cases, called \textbf{\emph{Gold}}, and confirm their reference ASGs. A large language model (LLM) agent constructs the training and validation cases, called \textbf{\emph{Silver}}, by checking CTI-described relations against PG evidence and repairing the retriever, parser, or script when evidence is missing (Sections~\ref{sec:gold-construction} and~\ref{sec:silver-construction}).

At investigation time, the first step takes the PG and its POI edges and returns a compact subgraph of potentially attack-related events (Challenge \emph{C4}). A text cross-encoder compares candidate events with the alert, while a directed graph model incorporates dependency direction, neighborhood structure, repetition, and novelty relative to benign activity. Their fused scores select the PG evidence passed to retrieval (Section~\ref{sec:locate-subgraph}).

The second step takes this subgraph and the KG collection and returns candidate article-level KGs for detailed alignment (Challenges \emph{C3} and \emph{C4}). Directly searching for partial, many-to-many edge correspondences across every KG would create a large search space. We therefore adopt a coarse-to-fine strategy inspired by local feature matching in computer vision~\cite{sarlin2020superglue,sun2021loftr}: first identify graphs containing behavior similar to the attack subgraph, then search for detailed edge correspondences only within the retrieved graphs. Intuitively, this resembles locating an object within an image before examining its constituent details. Adapting dual-softmax assignment from visual feature matching~\cite{sun2021loftr}, we aggregate mutually preferred node and edge correspondences into a relevance score for each KG; whitening limits domination by generic relations. Every KG above a validation-selected threshold is retained, allowing several articles to explain complementary attack stages; if none qualifies, the pipeline records insufficient CTI support. This retrieval step narrows the edge-matching search to relevant knowledge using \emph{graph semantics}, while the next step learns the \emph{security semantics} needed for accurate many-to-many alignment (Section~\ref{sec:locate-reports}).

The third step takes the recovered PG evidence and the retrieved KGs and returns explicit links between PG events and KG relations (Challenges \emph{C2} and \emph{C3}). We define a \emph{PG unit} as the set of PG edges with identical source--relationship--destination text, so repeated events form one PG unit, with links to the original events retained; alert events provide context and are not themselves matched. An LLM fine-tuned on Silver's correspondences maps each PG unit to a set of KG edges, learning which combinations of system events implement CTI-described attack steps. Empty outputs preserve unsupported behavior as unmatched, and voting over shuffled inputs retains consistent links. Finally, an ASG generator takes these links, their supporting PG events, and their CTI source sentences and produces a case-level report organizing the supported attack steps and their progression (Sections~\ref{sec:match-edges}--\ref{sec:generate-asg}).

Evaluation on 27 expert-verified cases with benign background activity and 713 CTI articles supports three findings. \textbf{First, explicit alignment produces more complete reports.} LODESTONE's reports recover 70\% of the reference attack facts, achieve an 82\% description score, and identify the attack in 91\% of cases, whereas the strongest evaluated baseline pipeline and general-purpose investigation agent achieve description scores of 47\% and 57\% and recover 38\% and 49\% of reference facts (Section~\ref{sec:results-rq1}). With the same report generator, supplying only the text of the articles retrieved for the located PG yields 72\% and 52\%: linking CTI knowledge to observed events outperforms supplying the retrieved articles alone. \textbf{Second, existing provenance and graph-matching methods do not bridge the PG--KG gap.} Provenance-reduction methods either return tens of thousands of edges per case or return compact subgraphs that miss most attack edges; LODESTONE recovers 89\% of annotated attack edges in subgraphs averaging 484 edges, whereas methods with similar coverage return 82 to 110 times as many (Section~\ref{sec:results-rq2}). Graph-matching methods designed for provenance graphs reach at most 56\% mean reciprocal rank when retrieving CTI KGs, whose vocabulary and topology differ from the PG's, below lexical BM25~\cite{robertson2009bm25} (89\%); LODESTONE reaches 98\%, and 95\% from its own located subgraphs (Section~\ref{sec:results-rq3}). \textbf{Third, the data engine autonomously turns the security knowledge in real-world CTI into training data.} It pairs each article's KG with PGs of actually executed behaviors and repairs its own tools when evidence is missing; trained only on these cases, the matcher gains KG-edge match recall at all five model sizes, from 52\% to 75\% at 4B (Section~\ref{sec:results-rq4}). These links make the supported interpretations traceable to concrete events and source passages, helping investigators verify the explanation and reconstruct the attack.

\section{Background and Threat Model}
\label{sec:problem-formulation}

\subsection{Cyber Attack Investigation}

\begin{figure*}[t]
    \centering
    \includegraphics[width=0.95\textwidth]{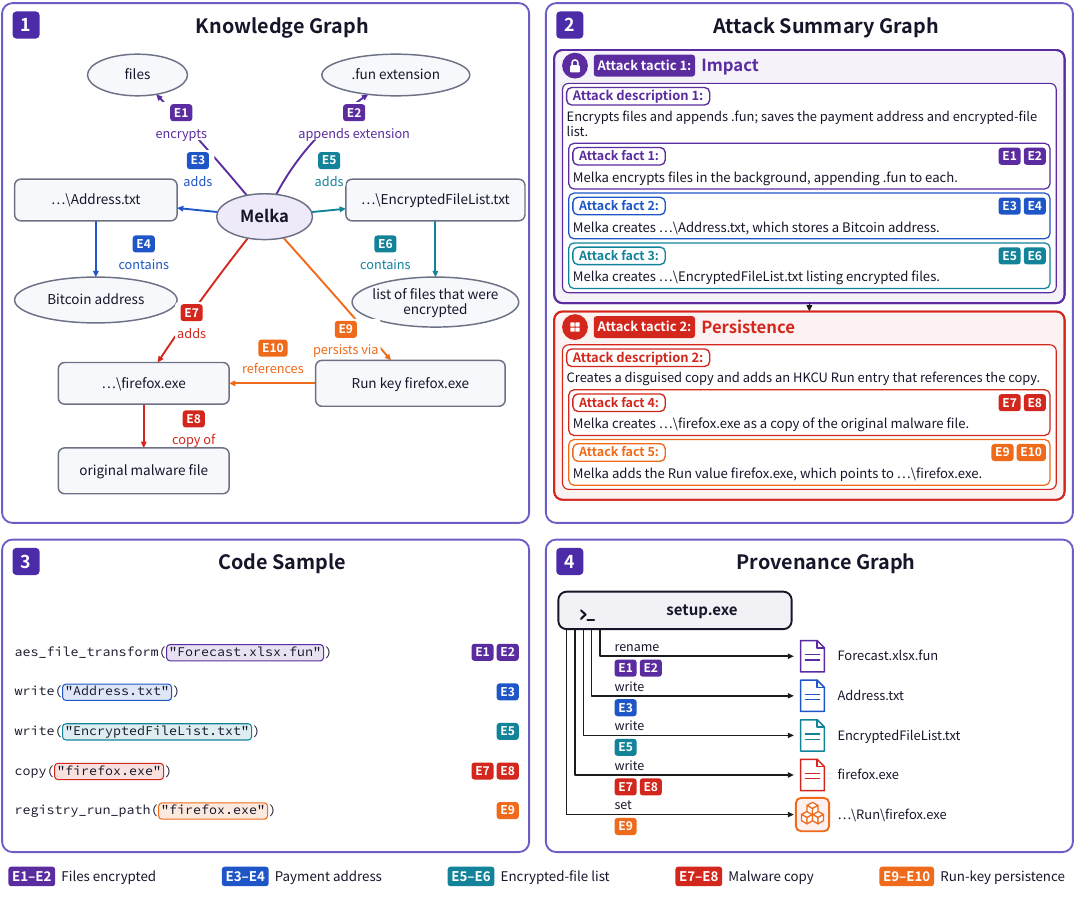}
    \caption{One attack in four forms: the CTI knowledge graph with its edges labeled E1--E10, the reference attack summary graph, the reproduction code, and the provenance graph.}
    \label{fig:benchmark-representation-mapping}
\end{figure*}

A \emph{cyber attack investigation} reconstructs an attack from observed system activity and explains its steps and goals. In our setting, the investigation starts from upstream alerts and has access to a host provenance graph (PG) and a knowledge base of CTI articles represented as knowledge graphs (KGs). It must identify the attack-related activity, find the articles that explain it, and establish which observed events support which reported behaviors. An alert may be a false alarm, and the knowledge base may contain no article that explains it. Figure~\ref{fig:benchmark-representation-mapping} illustrates the investigation concepts with a Melka ransomware case: the PG records file and registry operations, the KG supplies their attack meanings, and the reference \emph{attack summary graph} (ASG; Section~\ref{sec:story-graph}), a summary of the attack's goals and steps, organizes those meanings into impact and persistence stages; the code panel shows how the observable behavior is reproduced.

\subsection{Provenance Graph (PG)}
\label{sec:provenance-graph}

The PG records activity on the investigated host during an observation window. Following audit-based provenance representations~\cite{hossain2017sleuth}, its nodes represent system objects, such as processes, files, registry objects, and network endpoints, and its directed edges represent recorded system events. We call these \emph{PG event edges}, shortened to \emph{PG edges}. We define a \emph{PG unit} as the set of PG edges with identical source--operation--target text; it keeps links to the original events. In Figure~\ref{fig:benchmark-representation-mapping}, panel 4 shows a process, \texttt{setup.exe}, linked to file and registry objects by operations such as rename, write, and set. Our PG records operations on objects but omits their internal contents, such as document text or image data. The full graph records all events in the window: attack behavior and benign background activity (noise).

A point-of-interest (POI) edge corresponds to the system event that triggered an upstream alert. It anchors the investigation but does not establish that an attack has occurred. Starting from the POIs, the investigation seeks an \emph{attack-related PG subgraph}: a compact part of the full PG that contains the attack-relevant events and their dependencies.

\subsection{CTI Knowledge Graph (KG)}
\label{sec:cti-knowledge-graph}

Each CTI article is represented by an \emph{article-level KG}. Its nodes represent entities mentioned in the article, such as malware, software, files, and network endpoints, and its directed edges express the relationships stated between these entities~\cite{huang2024ctikg,li2022attackg}. We call these \emph{KG relation edges}, shortened to \emph{KG edges}. In Figure~\ref{fig:benchmark-representation-mapping}, panel 1 describes Melka encrypting files (E1), appending the \texttt{.fun} extension (E2), and persisting through a Run key (E9--E10), expressing attack behavior beyond the PG's operation vocabulary.

An article-level KG is relevant to an investigation when the behavior described in the article helps explain activity observed in the PG. The entities and KG edges relevant to the investigated attack form an \emph{attack-related KG subgraph}. An article may also describe other attacks or behaviors, so this subgraph may cover only part of its article-level KG. Some relations in the attack-related subgraph may have no observable PG evidence.

\subsection{KG--PG Matching}
\label{sec:kg-story-relationship}

KG--PG matching relates attack behavior described in CTI to observed system activity in a PG~\cite{milajerdi2019poirot}. In LODESTONE, it operates at two levels. \emph{Article-level KG retrieval} finds CTI articles relevant to a given attack-related PG subgraph~\cite{qiu2026aptcglp} and returns their article-level KGs. Article-level relevance is many-to-many: one PG subgraph may draw on several article-level KGs, and the same article-level KG may be relevant to PG subgraphs from different investigations. \emph{Edge-level matching} then links PG edges to KG edges within the retrieved graphs~\cite{milajerdi2019poirot}.

A \emph{KG--PG match} links a KG edge to the PG edges that evidence the behavior it describes. A single KG edge may require several PG edges: for example, downloading a file involves receiving data from a network endpoint and writing the file. Conversely, one PG edge may support more than one KG edge. In Figure~\ref{fig:benchmark-representation-mapping}, the rename event targeting \texttt{Forecast.xlsx.fun} in panel 4 supports both encryption (E1) and extension appending (E2) in panel 1, because in the real world Melka, a Jigsaw variant, gives each file the \texttt{.fun} extension after encrypting it~\cite{sonicwall2019melka,trendmicro2016jigsaw}. Edge-level matching is likewise many-to-many and partial: a PG edge may match zero, one, or several KG edges, and a KG edge may have no observable PG counterpart.

Each match retains the original PG event identifiers and the CTI source sentences associated with the KG edge. Adding these matches produces a \emph{KG-enhanced PG}: the original PG subgraph with its nodes, edges, and structure retained, and matched PG edges annotated with their KG edges and source sentences. PG edges without a match remain in the subgraph. The KG-enhanced PG is then converted into an investigation report (Section~\ref{sec:generate-asg}).

\subsection{Attack Summary Graph}
\label{sec:story-graph}

An \emph{attack summary graph} (ASG) is a reference representation of an attack's steps and goals, as illustrated in Figure~\ref{fig:benchmark-representation-mapping}. Each node represents a stage labeled with a MITRE ATT\&CK \emph{attack tactic}, the goal pursued by the steps in that stage. The same tactic may recur at different stages, represented by separate nodes. Directed edges indicate temporal order between stages. Panel 2 of Figure~\ref{fig:benchmark-representation-mapping} illustrates the hierarchy: the impact stage contains one attack description split into three facts about encryption, the payment-address file, and the encrypted-file list; the persistence stage contains a description split into two facts about the malware copy and its Run entry. Labels such as E1--E2 connect each fact to the KG relations it summarizes.

Under each tactic, an \emph{attack description} states one attack step. Each description is split into \emph{attack facts}, minimal statements that cannot be split further and can be judged individually. Gold's descriptions and facts are checked and confirmed by humans, and each Gold reference ASG serves as the ground truth of the attack that actually took place in its case, against which generated reports are evaluated. Silver's ASGs are generated by the agent during construction and are used neither by the pipeline nor for training. An ASG is independent of the KG-enhanced PG produced by automatic matching. Construction is described in Sections~\ref{sec:gold-construction} and~\ref{sec:silver-construction}.

\subsection{Threat Model}
\label{sec:threat-model}

We consider attackers who exploit system vulnerabilities to exfiltrate or manipulate sensitive data, potentially interleaving attack steps with benign activity. Because operating systems, other installed software, and threat variants differ, the indicators recorded in provenance logs can differ from those in CTI. We assume CTI is curated from sources widely used by analysts, with sampled manual inspection; incomplete knowledge remains possible. Deliberate CTI/KG poisoning and prompt injection are outside our scope; adversarial robustness is complementary~\cite{adversriallearning1,adversriallearning2,adversriallearning3}.

The kernel and deployed auditing framework, such as Linux Audit, Event Tracing for Windows (ETW), DTrace, or Sysdig~\cite{auditd,etw,dtrace,sysdig}, form the trusted computing base. We assume untampered logs and faithful PG construction, supported by trusted provenance capture and tamper-evident logging~\cite{trustkernel,tamperlog}. Implicit flows that bypass auditing, including side channels, are outside scope. Unrecorded behavior and object contents cannot be inferred from PG evidence alone: in Figure~\ref{fig:benchmark-representation-mapping}, panel 4 records a write to \texttt{Address.txt}, but does not establish its Bitcoin-address content (E4 in panel 1).

\FloatBarrier
\section{Approach}
\label{sec:approach}

\begin{figure*}[t]
    \centering
    \includegraphics[width=\textwidth]{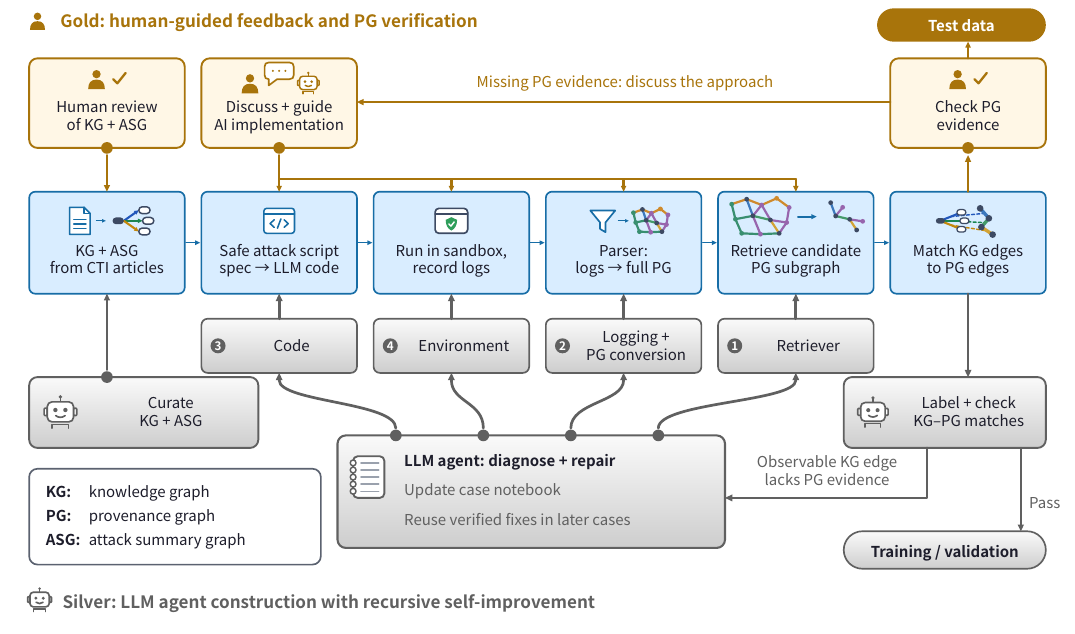}
    \caption{How Gold and Silver cases are built: humans curate Gold, and an LLM agent builds Silver and repairs its tools whenever PG evidence is missing.}
    \label{fig:dataset-tiers}
\end{figure*}

\begin{figure*}[t]
    \centering
    \includegraphics[width=\textwidth]{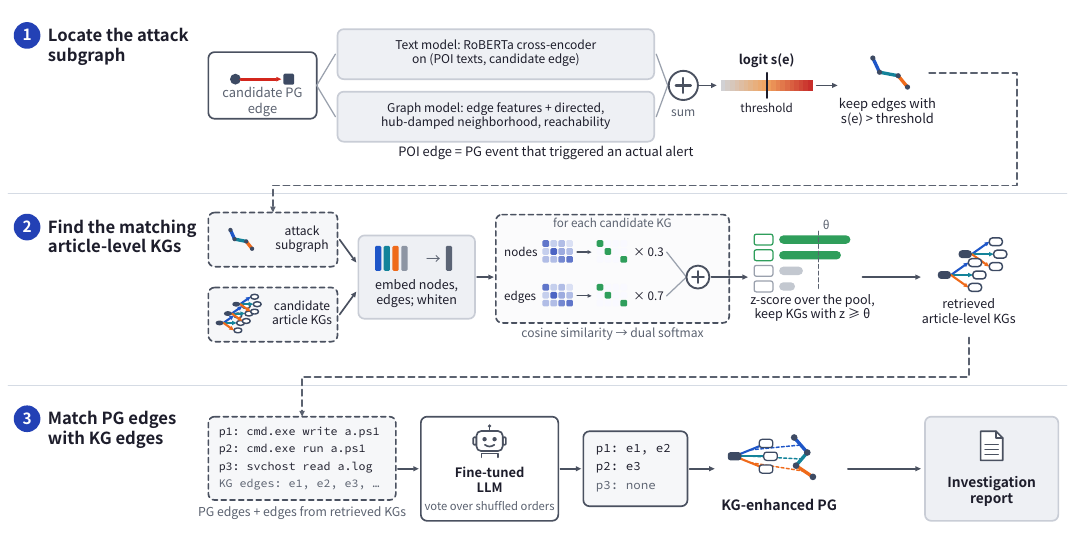}
    \caption{The three steps of an investigation: locate the attack subgraph, retrieve the matching article KGs, and match PG edges to KG edges.}
    \label{fig:pipeline-steps}
\end{figure*}

LODESTONE has a two-tier data engine (Figure~\ref{fig:dataset-tiers}) and a three-step investigation pipeline (Figure~\ref{fig:pipeline-steps}). The engine builds human-verified \textbf{Gold} cases for testing and agent-built \textbf{Silver} cases for training and validation (C1--C2); from POI edges, the pipeline locates the attack-related PG subgraph, retrieves article-level KGs, and matches PG edges to KG edges (C2--C4). All of its thresholds, weights, and other settings are selected on Silver validation cases, never on Gold (Section~\ref{sec:data-splits}).

\subsection{Gold: Human-Verified Benchmark}
\label{sec:gold-construction}

\noindent\textbf{Threat selection and KG subgraph.} An LLM converts CTI articles into article-level KGs. A curator selects a coherent installation or runtime chain for one threat (e.g., in PlugX, a dropper runs \texttt{esetservice.exe}, which loads \texttt{http\_dll.dll}, which reads \texttt{lang.dat}), checks its KG edges against the source passages, corrects errors, adds supported omissions, and removes unsupported or out-of-scope relations to form the attack-related KG subgraph.

\noindent\textbf{Reference ASG construction.} An LLM drafts attack descriptions and atomic facts from the confirmed KG subgraph; the curator corrects and confirms both against KG edges and CTI passages. Descriptions carry ATT\&CK techniques and form temporally linked, tactic-labeled stages. The confirmed ASG (Section~\ref{sec:story-graph}, Table~\ref{tab:asg-layers}) is the reference for report evaluation (Section~\ref{sec:metrics}).

\begin{table*}[t]
\centering
\caption{The three layers of the reference ASG, illustrated by the ransomware investigation in Figure~\ref{fig:benchmark-representation-mapping}.}
\label{tab:asg-layers}
\footnotesize
\setlength{\tabcolsep}{4pt}
\begin{tabular}{@{}p{0.11\textwidth}p{0.30\textwidth}p{0.38\textwidth}p{0.15\textwidth}@{}}
\toprule
Layer & Definition & Example & Used by \\
\midrule
Attack tactic & Stage goal labeled with an ATT\&CK tactic; tactics may recur. & Persistence (TA0003) & Organizes the ASG (not scored) \\
Attack description & Human-confirmed attack step summarizing one or more facts, with its ATT\&CK technique. & ``Creates a disguised copy and an HKCU Run entry referencing it.'' (T1547.001) & Description score \\
Attack fact & Human-confirmed minimal claim abstracted from one or more KG edges, independently assessable. & ``Melka adds Run entry \texttt{firefox.exe}, pointing to \texttt{\%APPDATA\%\textbackslash Frfx\textbackslash firefox.exe}.'' & Fact recall \\
\bottomrule
\end{tabular}
\end{table*}

In Figure~\ref{fig:benchmark-representation-mapping}, ten KG edges support five attack facts and five code calls produce five PG events; E4 and E6 describe data stored inside files, and E10 the data of a registry value; the PG records interactions between system entities, not the data they store, so these edges have no PG matches.

\noindent\textbf{CTI-grounded scripts and execution.} Our \emph{Safely-implement} design (Appendix~\ref{app:safe-implement}) specifies the entities, observable operations, prerequisites, and reversible benign surrogates for the KG behaviors. Guided by the curator, the LLM agent writes the code, and the curator validates the resulting PG evidence. The four code roles (Table~\ref{tab:code-kinds}) run with background-noise code in the VM, while the host manages the environment, virtual network, and log storage. Capture begins 30--90\,s before the core and ends 30\,s after it, so the PG also records background activity (Section~\ref{sec:candidate-pool}); after clean-up, the host retrieves the logs to build the full PG (Appendix~\ref{app:pg-construction}).

Behaviors that cannot be both safe and faithful are withheld with recorded reasons, and implemented behaviors may have unobservable details; neither changes CTI facts. Reported domains resolve to sandbox addresses, and each KG process entity is played by one process and its descendants. Appendix~\ref{app:safe-implement} gives the safety and consistency rules; Appendix~\ref{app:ethics} discusses ethics.

\begin{table}[t]
\centering
\footnotesize
\caption{Code roles and their place in a reproduced case.}
\label{tab:code-kinds}
\begin{tabular}{@{}p{0.22\columnwidth}p{0.71\columnwidth}@{}}
\toprule
Code & Role and execution phase \\
\midrule
Preparation & Before capture in the case VM: stage prerequisites and canary (decoy) inputs, without attack products. \\
Core & During capture in the case VM: perform specified behaviors alongside background activity. \\
Clean-up & After capture: revert case changes and verify the VM is ready for the next case. \\
Responder & Outside the case VM: provide the sandbox's remote services. \\
\bottomrule
\end{tabular}
\end{table}

\noindent\textbf{KG--PG correspondence annotation.} A \emph{candidate retriever}, written from the KG subgraph and code before execution, narrows more than a hundred thousand PG edges to a few hundred by names, paths, operations, and event chains, tolerating run-specific values such as process IDs, so that the agent searches a small space for each KG edge's PG edges. If an expected edge is missing, the agent first searches the full PG; if the edge is there, only the retriever is repaired.

The agent proposes correspondences, and the curator confirms the PG edges, possibly several or repeated, that faithfully show each KG behavior: creating a file requires its write, downloading requires receiving data and writing the file, and preparatory operations such as opening a file are not counted as matches; the PG edges confirmed this way are the \emph{attack edges} (Appendix~\ref{app:correspondence}).

\noindent\textbf{PG-based verification and feedback.} When a confirmed KG behavior lacks PG evidence, the agent, guided by the curator, checks the four components in Table~\ref{tab:missing-pg-evidence}, repairs the one at fault, and searches again, reconverts retained logs, or re-runs the case; the curator then rechecks the PG. Relations with confirmed safety or observability limits keep their CTI support and reasons without forced PG matches and are excluded from edge-matching scoring (Section~\ref{sec:metrics}). Section~\ref{sec:curation-records} describes the records and shared guidance.

\begin{table}[t]
\centering
\footnotesize
\caption{Possible sources of missing PG evidence.}
\label{tab:missing-pg-evidence}
\begin{tabular}{@{}p{0.31\columnwidth}p{0.62\columnwidth}@{}}
\toprule
Component & Possible cause \\
\midrule
Candidate retriever & Existing PG evidence is missed. \\
Logging and PG conversion & The event is not recorded by the telemetry or is lost during PG conversion. \\
Reproduction code & The behavior or its entities are implemented incorrectly. \\
Experimental environment & Configuration, permissions, or dependencies prevent execution or observation. \\
\bottomrule
\end{tabular}
\end{table}

\subsection{Silver: Agent Construction with Recursive Self-Improvement}
\label{sec:silver-construction}

Silver follows the Gold workflow (Figure~\ref{fig:dataset-tiers}), but an LLM agent, using sanitized guidance, curates the KG subgraph and ASG, implements the behaviors, and labels correspondences from execution evidence; KG edges retain their source passages.

\noindent\textbf{Recursive self-improvement.} Silver automates Gold's feedback loop: missing evidence triggers a correspondence recheck, diagnosis among the same four causes, repair, and renewed verification. In this \emph{recursive self-improvement} (RSI), verified tool repairs remain in the pipeline, and their causes, remedies, and conditions enter the case notebook. The KG subgraph, ASG, specification, code, and retriever are fixed within a round; revisions apply to the next.

\noindent\textbf{Case acceptance and data split.} Accepted cases need source-supported KG and ASG annotations, reproducibility, and PG evidence for every safe, faithful, observable behavior; other relations keep documented reasons. Silver is split by case after excluding overlaps with Gold (Section~\ref{sec:data-splits}).

\noindent\textbf{Notebooks and shared guidance.} \label{sec:curation-records}
Records make every label traceable to the decision and evidence behind it, and carry rulings from one case to the next (Table~\ref{tab:records}). Generalizable Gold rulings enter a shared guide, stripped of Gold entities, behavior chains, and answers before it guides Silver. Because agents sometimes skip or misread guidance, we wrap the agent in a hash-based \emph{harness}, the code around an agent that constrains its actions: every rule section carries a content hash. Before each VM run, annotation, or acceptance, the agent submits a reading record listing each required section by hash, with a quote and its application to the case; the entry point rejects a missing section or an outdated hash, and each correspondence must cite an applied rule.

\begin{table}[t]
\centering
\footnotesize
\caption{Curation records of Gold and Silver.}
\label{tab:records}
\begin{tabular}{@{}p{0.15\columnwidth}p{0.37\columnwidth}p{0.38\columnwidth}@{}}
\toprule
 & Gold & Silver \tabularnewline
\midrule
\raggedright Guide & \raggedright Rules from curator rulings. & \raggedright Gold's undisputed rules, no Gold content. \tabularnewline
\raggedright Case notebook & \raggedright Curator rulings, verbatim, per component. & \raggedright Agent decisions; runs, failures, and repairs per round. \tabularnewline
\raggedright Status & \raggedright Curator-confirmed components. & \raggedright Agent-checked components. \tabularnewline
\raggedright Purpose & \raggedright Trace labels to human rulings; reuse them. & \raggedright Keep verified repairs; expose regressions. \tabularnewline
\bottomrule
\end{tabular}
\end{table}

\subsection{Locating the Attack Subgraph from POI Edges}
\label{sec:locate-subgraph}

Every edge of the full PG is a candidate. The text branch, a RoBERTa cross-encoder~\cite{liu2019roberta}, compares each candidate's source--operation--target text with the concatenated distinct texts of the POI edges. The graph branch uses edge embeddings, endpoint degrees, event repetition, novelty relative to benign activity, and hop distance and directional reachability to the nearest POI endpoint; learned incoming and outgoing neighborhood summaries preserve direction, and inverse event-frequency weighting damps hubs.

Both branches are trained on Silver; their logits are summed, and edges above a fixed threshold form the located subgraph. Each seed uses the highest threshold whose mean output on Silver validation cases with attack edges reaches five times their mean labeled attack-edge count.

\subsection{Finding the Matching Article-Level KGs}
\label{sec:locate-reports}

\begin{figure}[t]
    \centering
    \includegraphics[width=\columnwidth]{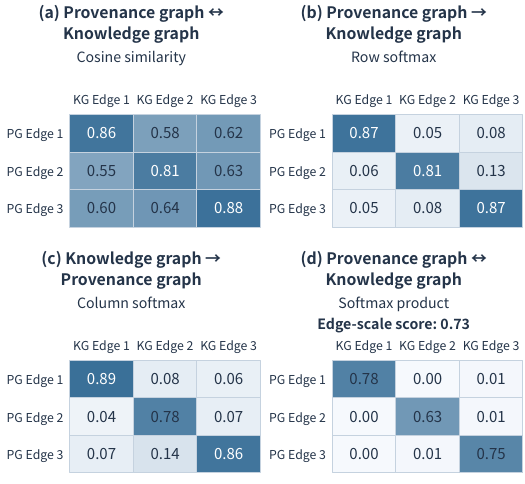}
    \caption{A dual-softmax example: (b) normalizes each row of (a), (c) normalizes each column, and (d) multiplies the two.}
    \label{fig:dual-softmax-example}
\end{figure}

The second step retrieves article-level KGs that can explain the located PG subgraph. Searching for partial, many-to-many correspondences against every edge of every KG would expose the matcher to a large candidate space. Inspired by local feature matching in computer vision~\cite{sarlin2020superglue}, in particular its coarse-to-fine form~\cite{sun2021loftr}, we first score candidate graphs for behavior similar to the attack subgraph and then restrict edge matching to the retrieved KGs, separating inexpensive graph-level retrieval from security-aware alignment of event combinations. Frozen embeddings represent node names and types and edge descriptions. A regularized whitening transform~\cite{su2021whitening}, fitted on the knowledge-base node and edge embeddings separately and applied to both PG and KG, reduces the dominance of shared components before cosine comparison. For nodes and for edges, dual softmax~\cite{sun2021loftr} combines row- and column-normalized preferences, rewarding pairs that prefer each other among competitors (Figure~\ref{fig:dual-softmax-example}); the graph score weights node and edge evidence 30\% and 70\%.

Scores are standardized across all candidate KGs of a query, and every KG above a threshold is retained, so several articles can explain complementary stages; their edges are the candidates for edge-level matching.

\subsection{Matching PG Units with Edges of the Retrieved KGs}
\label{sec:match-edges}

The third step poses edge-level matching as selection from the retrieved KGs' edges~\cite{typesafe2026jev,cloudflare2026clef}. Matching operates on PG units (Section~\ref{sec:provenance-graph}). For each PG unit, the matcher selects the possibly empty set of KG edges whose behaviors it supports; one PG unit may support several KG edges, and several PG units the same edge.

A 4B LLM is fine-tuned with LoRA~\cite{hu2022lora} on the PG-unit-to-KG-edge labels of Silver training records, revised once by an LLM outside the sandbox (Appendix~\ref{app:genai}). At inference, it answers the original input and seven copies with PG units and options shuffled, keeping links with at least four votes; these annotate the original PG edges to form the KG-enhanced PG, with unmatched edges preserved. An optional fine-tuned 4B reviewer, the \emph{stepwise review}, revisits links with adjacent PG edges and can remove benign PG units (Section~\ref{sec:results-rq4}).

\subsection{Generating the Investigation Report}
\label{sec:generate-asg}

The \emph{report generator} is an LLM without task-specific training: report quality can be graded only by an LLM judge calibrated against human grading (Section~\ref{sec:metrics}), which gives no reward reliable enough for reinforcement learning or other post-training. It receives only linked KG edges, their CTI source sentences, and the supporting PG units; if retrieval returns no KG, this evidence is empty and the report records a no-match outcome. It describes supported attack steps in temporal order, one atomic fact per step (one subject, one action, and one object), cites their CTI evidence, and omits unsupported steps; unmatched PG edges remain in the KG-enhanced PG. LODESTONE's trained components are thus the two step-1 branches and the step-3 matcher.

\section{Evaluation}
\label{sec:evaluation}

\label{sec:research-questions}
We evaluate LODESTONE through four research questions.

\noindent\textbf{RQ1 (Report quality).} How closely do the generated investigation reports match the reference attack descriptions and facts?

\noindent\textbf{RQ2 (Localization).} How well does the located PG subgraph cover attack edges while limiting its size?

\noindent\textbf{RQ3 (Retrieval).} How accurately does LODESTONE retrieve the source articles' KGs?

\noindent\textbf{RQ4 (Edge matching).} How accurately does LODESTONE find correspondences between PG edges and KG edges?

\subsection{Setup}
\label{sec:eval-conditions}

\subsubsection{Data Splits}
\label{sec:data-splits}
Gold's 27 cases (24 Windows, 3 Linux) are used only for evaluation; a case averages 270{,}366 PG edges, of which 159 are attack edges, and 2.1 source articles. The subgraph-localization models (Section~\ref{sec:locate-subgraph}) are trained and validated on 78 Silver cases (55 and 23 per seed); the matcher (Section~\ref{sec:match-edges}) is trained on 186 Silver records from 106 cases. Silver cases sharing a source article or malware family with Gold are excluded, and the rest are split by case into training and validation (70\%/30\%); checkpoints and hyperparameters such as thresholds, temperatures, and fusion weights are selected on validation.

\noindent\emph{POIs.} The first POI is the latest attack event with an observable effect, i.e., a state change: writing, creating, renaming, or deleting a file, starting a process, sending network data, or setting a registry value; reading or opening an object, or receiving network data, does not count. Further POIs come from different attack steps, in decreasing order of an LLM's anomaly rating. A benign investigation (RQ1's false-explanation rate) takes as its single POI the latest such event in the background window recorded before the attack.

\subsubsection{Noise}
\label{sec:candidate-pool}
\noindent\emph{Background activity.} Benign tasks run in the same VM before, during, and after the reproduced behavior (Appendix~\ref{app:background-activity}).

\noindent\emph{Knowledge base.} End-to-end evaluation uses the 1:10 candidate pool of 713 article-level KGs, including source articles and distractors from PRIMUS~\cite{yu2025primus,primusseed_hf,primusfineweb_hf}. RQ3 additionally uses the 1:100 and 1:500 pools, containing 6{,}363 and 31{,}563 KGs, respectively.

\subsubsection{Metrics}
\label{sec:metrics}
\noindent\emph{RQ1.} Against the human-edited reference ASG, an LLM judge grades each attack description in a report as correct (2), partial (1), or absent (0); the \emph{description score} is the mean grade as a percentage of the maximum. Inspired by atomic-fact evaluation~\cite{min2023factscore,wei2024longform}, we define \emph{reference fact recall} (``fact recall'') as the fraction of reference attack facts stated, under a one-to-one matching between reference facts and report claims. Every method is instructed to state one fact per claim (Section~\ref{sec:generate-asg}); each claim is therefore credited for at most one fact, and a claim that bundles several facts is credited once. \emph{Attack identification} is the fraction of attack cases named correctly or by a recognized alias, not by a generic label such as ransomware. The \emph{false-explanation rate} is the fraction of benign-investigation reports that explain an attack; with source articles removed, the \emph{insufficient-support rate} is the fraction of reports leaving the attack unnamed. Non-agent methods share a fixed report generator (Section~\ref{sec:generate-asg}).

On 100 randomly sampled items (50 descriptions and 50 facts), GLM-5.3-Flash and Claude Sonnet 5.5 agree with the original DeepSeek-V4.1-Flash (DS) judge on 92\% and 98\% of description grades ($\kappa$ 0.92 and 0.98, linearly weighted for the three grade levels), respectively, and on 94\% and 94\% of fact labels (Cohen's $\kappa$ 0.88 and 0.88; one GLM timeout). Three authors (A--C) with security expertise, reviewing the same items with the judge's output shown, agree with 90\%, 88\%, and 88\% of its description grades (weighted $\kappa$ 0.89, 0.87, and 0.87) and 98\%, 96\%, and 88\% of its fact labels ($\kappa$ 0.96, 0.92, and 0.76); pairwise, the authors agree with each other on 84--92\% of description grades (weighted $\kappa$ 0.83--0.89) and 86--98\% of fact labels ($\kappa$ 0.72--0.96; all pairs in Table~\ref{tab:judge-agreement}, Appendix~\ref{app:judge-agreement}). Re-grading every report of Table~\ref{tab:rq1} with GLM-5.3-Flash in place of DeepSeek-V4.1-Flash as the judge leaves the ranking by description score unchanged (92\% agreement on description grades, 96\% on fact labels).

\noindent\emph{RQ2.} We report returned PG edges and \emph{PG attack-edge recall}, also called attack-edge coverage: the fraction of all human-annotated attack edges in the PG that are returned. Causality-based outputs are capped at 25\% of the PG (25 to 100{,}000 edges).

\noindent\emph{RQ3.} Queries use human-verified attack edges. Mean reciprocal rank (MRR) uses the highest-ranked source article; Hits@$k$ is the fraction of cases retrieving a source article among the top $k$ KGs; mean average precision (MAP) treats every source article of a case as relevant. MRR$^\dagger$ instead uses LODESTONE's step-1 output. For cases with $n$ source articles, \emph{article-level recall} is their fraction retrieved among the top $n$ KGs.

\noindent\emph{RQ4.} All methods match the same PG units (PG edges grouped by source--operation--target text; Section~\ref{sec:provenance-graph}) from LODESTONE's first step against the KGs from its second step, judged by the human-verified annotations. \emph{KG-edge match recall} is the fraction of scored KG edges correctly matched to at least one PG unit. \emph{Linked PG units} counts PG units linked to any KG edge; at equal recall, fewer is better. Scoring omits PG units with the same text as a POI edge, and KG edges without remaining verified PG evidence or marked not implementable: human review found that 79 of the 409 Gold KG edges (19\%) have no safe and faithful counterpart in the telemetry, such as Dridex injecting code into the system process \texttt{explorer.exe}, which we did not reproduce for safety, and Jaqw's fake Windows-update alert, an on-screen display that leaves no trace in the telemetry (Table~\ref{tab:not-implementable}).

A KG edge is scored when it belongs to a retrieved KG and keeps verified PG evidence in at least one step-1 PG unit; edges of KGs that the second step did not retrieve are outside the scoring set. Scored edges without a prediction count as missed.

One exemption, the \emph{PG-only exemption}, covers steps that the PG-only reports of RQ1 (the generator given only the located PG) already describe: if every attack description supported by a KG edge receives full RQ1 grades in every seed of these reports, the edge leaves a method's denominator unless that method matches it correctly. Denominators can therefore differ between methods; cases without scored KG edges are omitted. All step metrics are macro-averaged over cases. Because the exemption only removes edges that a method fails to match, it can only raise recall.

\subsubsection{Baselines}
\label{sec:baselines}
Table~\ref{tab:training-data} groups compared methods by their training sources before investigation. Attack cases can include benign background events.

\begin{table}[t]
\centering
\caption{Sources of task-specific training data.}
\label{tab:training-data}
\footnotesize
\begin{tabular}{@{}p{0.23\columnwidth}p{0.70\columnwidth}@{}}
\toprule
Training source & Methods\tabularnewline
\midrule
None & \raggedright BackTracker~\cite{king2003backtracker}, SLEUTH~\cite{hossain2017sleuth}, POIROT~\cite{milajerdi2019poirot}, DEPIMPACT~\cite{fang2022depimpact}, SPARSE~\cite{ying2026sparse}, DEPCOMM~\cite{xu2022depcomm}, BM25~\cite{robertson2009bm25}, query likelihood~\cite{zhai2001smoothing}, frozen BGE-M3~\cite{chen2024bgem3}, dual-softmax assignment~\cite{sun2021loftr}, ThreatRaptor~\cite{gao2021threatraptor}, FGW~\cite{vayer2019fgw}, zero-shot GLM-5.3~\cite{glm5team2026glm5,zai2026glm53_hf} and DeepSeek V4.1 Flash~\cite{deepseekai2026deepseekv41flash}, SWE-agent~\cite{yang2024sweagent} and Codex~\cite{openai2025codexcli}\tabularnewline
\raggedright Benign windows & \raggedright PrioTracker~\cite{liu2018priotracker}, ORTHRUS~\cite{jiang2025orthrus}, DeepHunter~\cite{wei2021deephunter}, MEGR-APT~\cite{aly2024megrapt}, NeuroMatch~\cite{ying2020neuromatch}, ProvG-Searcher~\cite{altinisik2023provgsearcher}, APT-CGLP~\cite{qiu2026aptcglp}\tabularnewline
\raggedright Attack cases & \raggedright ATLAS~\cite{alsaheel2021atlas}, Sentence-BERT~\cite{reimers2019sbert}, BERT~\cite{devlin2019bert}, RoBERTa~\cite{liu2019roberta}, and DeBERTa~\cite{he2023debertav3} pair classifiers\tabularnewline
\raggedright Benign windows + attack cases & \raggedright TREC~\cite{lv2024trec}\tabularnewline
\bottomrule
\end{tabular}
\end{table}

\subsection{RQ1: Report Quality}
\label{sec:results-rq1}

We compare LODESTONE with baseline three-step pipelines and with general-purpose agents that investigate directly from the PG, POIs, and articles.

\begin{table}[t]
\centering
\caption{Report quality on Gold attack cases.}
\label{tab:rq1}
\footnotesize
\setlength{\tabcolsep}{2.8pt}
\begin{tabular}{@{}lrrr@{}}
\toprule
Evidence or method & \begin{tabular}[b]{@{}c@{}}Description\\score\end{tabular} & \begin{tabular}[b]{@{}c@{}}Fact\\recall\end{tabular} & \begin{tabular}[b]{@{}c@{}}Attack\\identified\end{tabular} \\
\midrule
LODESTONE & \textbf{82\%} & \textbf{70\%} & \textbf{91\%} \\
\quad with additional stepwise review & 76\% & 66\% & 87\% \\
\midrule
\multicolumn{4}{@{}l}{\emph{Evidence ablations}}\\
PG only & 73\% & 62\% & 32\% \\
Retrieved article text & 72\% & 52\% & \textbf{91\%} \\
\midrule
\multicolumn{4}{@{}l}{\emph{Component ablations (one component replaced)}}\\
Locate: PrioTracker & 37\% & 29\% & 76\% \\
Retrieve: BM25 & 68\% & 58\% & 74\% \\
Match: BERT (Silver-trained) & 80\% & 67\% & \textbf{91\%} \\
Match: untuned Qwen3.5-4B & 51\% & 41\% & 82\% \\
Match: FGW & 71\% & 61\% & 89\% \\
\midrule
\multicolumn{4}{@{}l}{\emph{Baseline three-step pipelines}}\\
PrioTracker + BM25 + BERT & 28\% & 20\% & 57\% \\
PrioTracker + BM25 + GLM-5.3 & 47\% & 38\% & 74\% \\
\midrule
\multicolumn{4}{@{}l}{\emph{General-purpose agents}}\\
GLM-5.3 + SWE-agent & 57\% & 49\% & 63\% \\
Qwen3.5-4B + SWE-agent & 38\% & 34\% & 56\% \\
Qwen3.5-4B + Codex & 41\% & 31\% & 41\% \\
Qwen3.5-0.8B + SWE-agent & 0\% & 0\% & 0\% \\
Qwen3.5-0.8B + Codex & 0\% & 0\% & 0\% \\
\bottomrule
\end{tabular}
\end{table}

Results average two seeds, except the GLM-5.3 pipeline, which uses one report per case because the 753B-parameter model~\cite{zai2026glm53_hf} is costly and slow.

\subsubsection{Comparison with Baseline Pipelines}
\label{sec:rq1-baseline-pipelines}
The baseline pipelines replace LODESTONE's steps with the strongest alternatives: PrioTracker, the highest-coverage other locator under the capped RQ2 protocol; BM25, the best retrieval baseline in the 713-article pool; and BERT, the best non-LLM matcher in RQ4, or a zero-shot GLM-5.3 matcher. They share LODESTONE's report generator, and PrioTracker's edge budget is set on Silver by LODESTONE's rule.

LODESTONE reaches an 82\% description score, 70\% fact recall, and 91\% attack identification (Table~\ref{tab:rq1}); the BERT and GLM-5.3 pipelines reach description scores of 28\% and 47\%. Replacing one LODESTONE component at a time lowers the description score most when localization is replaced (37\%), and to 68\% when retrieval is replaced. Replacing the matcher also lowers it, to 51\% with the untuned 4B model and 71\% with FGW; a matcher trained on Silver holds up better, but even the BERT classifier trained on Silver reaches only 80\%, below LODESTONE's 82\%. Evidence ablations keep LODESTONE's report generator and change only its evidence: the located PG alone yields a 73\% description score, and the full text of the articles retrieved for that PG, without the PG itself, yields 72\%. Stepwise review lowers the description score to 76\%, and fact recall and attack identification also decline.

\subsubsection{Direct Investigation with General-Purpose Agents}
\label{sec:rq1-direct-agents}
Agents receive the full noisy PG, its POIs, and all 713 article texts in an offline sandbox, search on their own, and write reports in the same format without LODESTONE's intermediate results. We run SWE-agent with GLM-5.3, Qwen3.5-4B~\cite{qwenteam2026qwen35}, or Qwen3.5-0.8B, and Codex with the two Qwen models, with a budget of 80 model calls per case; reports use LODESTONE's format, and a missing report counts as empty.

The best agent in Table~\ref{tab:rq1}, GLM-5.3 with SWE-agent, reaches a 57\% description score, against LODESTONE's 82\%, and submits no report in 16 of 54 runs, all of which stop within the first few steps on a context-window or tool-call format error. Qwen3.5-4B reaches 38\% with SWE-agent and 41\% with Codex; the 0.8B configurations almost never submit a report.

In 22 benign investigations, LODESTONE wrongly explains an attack in 15 of 44 reports (34\%), against 36--86\% for agents that produce reports; with source articles removed, 89\% of its reports leave the attack unnamed, against 76--85\%. With the optional stepwise review (Section~\ref{sec:match-edges}), these rates are 32\% and 87\%.

\noindent\textbf{Case Study.} In MedusaLocker, all three methods identify the ransomware and report the \texttt{EnableLinked\-Connections} registry setting and scheduled-task persistence. LODESTONE's two reports also recover the file search, encrypted-file extensions, and ransom notes; the best pipeline omits the extensions and notes, and the best agent describes the search indirectly in one run and omits it in the other (Table~\ref{tab:rq1-case}).

\begin{table}[t]
\centering
\caption{Report quality in the MedusaLocker case.}
\label{tab:rq1-case}
\footnotesize
\setlength{\tabcolsep}{3pt}
\begin{tabular}{@{}lrr@{}}
\toprule
Method & \begin{tabular}[b]{@{}c@{}}Description\\score\end{tabular} & \begin{tabular}[b]{@{}c@{}}Fact\\recall\end{tabular} \\
\midrule
LODESTONE & \textbf{87.5\%} & \textbf{85.7\%} \\
PrioTracker + BM25 + GLM-5.3 & 62.5\% & 57.1\% \\
GLM-5.3 + SWE-agent & 68.8\% & 71.4\% \\
\bottomrule
\end{tabular}
\end{table}

\noindent\textbf{Observability Limits.} In Dridex, the reference ASG includes content-dependent facts: the Word attachment contains macro code, an embedded ActiveMime object holds an encrypted OLE file, and the received configuration lists banking websites. The PG's process starts, network transfers, and file accesses cannot reveal these contents.

\subsection{RQ2: Locating the Attack Subgraph}
\label{sec:results-rq2}

\begin{table}[t]
\centering
\caption{Attack-subgraph localization on Gold with up to five POIs.}
\label{tab:rq2}
\footnotesize
\setlength{\tabcolsep}{4pt}
\begin{tabular}{@{}lrr@{}}
\toprule
Method & Edges/case $\downarrow$ & \begin{tabular}[b]{@{}c@{}}PG attack-\\edge recall $\uparrow$\end{tabular} \\
\midrule
LODESTONE & 484 & 89\% \\
\quad Text model only & 478 & 81\% \\
\quad Graph model only & 520 & 86\% \\
\midrule
\multicolumn{3}{@{}l}{\emph{Other text models}}\\
BGE-M3, frozen & 401 & 47\% \\
\quad with graph model & 508 & 88\% \\
Qwen3.5-0.8B & 450 & 86\% \\
\quad with graph model & 318 & 89\% \\
Qwen3.5-4B & 314 & 86\% \\
\quad with graph model & 392 & 93\% \\
\midrule
BackTracker~\cite{king2003backtracker} & 22{,}062 & 78\% \\
PrioTracker~\cite{liu2018priotracker} & 53{,}015 & 93\% \\
SLEUTH~\cite{hossain2017sleuth} & 27 & 18\% \\
DEPIMPACT~\cite{fang2022depimpact} & 19{,}024 & 60\% \\
SPARSE~\cite{ying2026sparse} & 4{,}463 & 49\% \\
ORTHRUS~\cite{jiang2025orthrus} & 39{,}574 & 90\% \\
DEPCOMM~\cite{xu2022depcomm} & 1{,}476 & 20\% \\
\midrule
ATLAS~\cite{alsaheel2021atlas} & 20{,}838 & 85\% \\
TREC~\cite{lv2024trec} & 32{,}362 & 77\% \\
\bottomrule
\end{tabular}
\end{table}

\begin{figure*}[t]
\centering
\includegraphics[width=\textwidth]{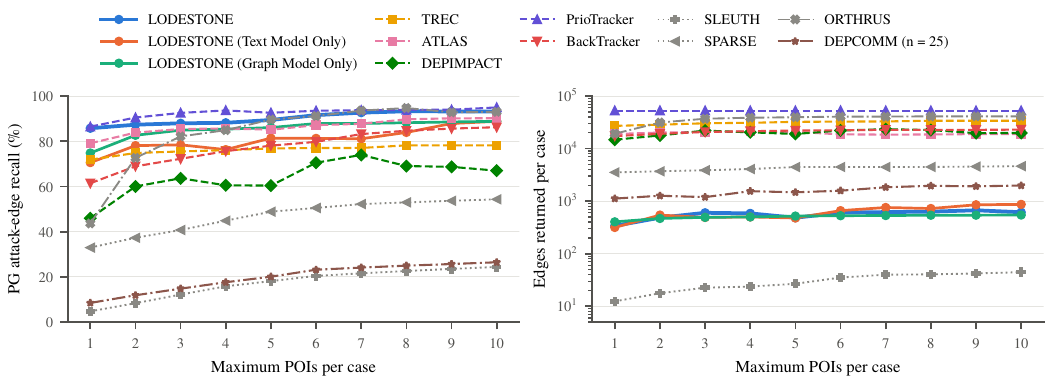}
\caption{How localization changes as more POIs are given per case: PG attack-edge recall (left) and returned edges (right).}
\label{fig:rq2-poi}
\end{figure*}

With up to five POIs, LODESTONE covers 89\% of the annotated attack edges while returning 484 PG edges per case (Table~\ref{tab:rq2}). ORTHRUS reaches similar recall (90\%) with about 82 times as many edges, and PrioTracker reaches 93\% with about 110 times as many, hitting the cap on causal outputs (Section~\ref{sec:metrics}) in every case. Causality-based trackers (BackTracker, PrioTracker) follow system dependencies from the POIs, and routine activity often shares those dependencies, so their outputs grow large. Weighting and pruning dependencies (DEPIMPACT) yields a somewhat smaller output but recovers only 60\% of the annotated attack edges. Methods that build a compact attack scenario or summary (SLEUTH, DEPCOMM) aim at a concise picture rather than complete evidence, and each covers about a fifth of the attack edges. Learning-based methods (ORTHRUS, ATLAS, TREC) reach 77--90\% recall but return about 21{,}000 to 40{,}000 edges per case. LODESTONE instead scores each edge against the POIs' text and graph context, so its output concentrates on attack-related edges.

At comparable output sizes, fusion adds eight percentage points of recall over the text branch alone and three over the graph branch alone, and the graph branch helps each of the three alternative text encoders. A fine-tuned LLM can also serve as the text encoder: Qwen3.5-4B with the graph branch reaches the highest Gold recall (93\% with 392 edges). Choosing by Gold would leak the test set into model selection, so downstream experiments keep RoBERTa, which has the highest average precision on Silver validation.

Raising the POI limit from one to ten increases recall from 86\% to 93\%, while returned subgraphs range from 339 to 671 edges per case (Figure~\ref{fig:rq2-poi}). Cases with fewer available POIs contribute their full POI sets.

\subsection{RQ3: Finding the Matching Article-Level KGs}
\label{sec:results-rq3}

\begin{table}[t]
\centering
\caption{Article-level KG retrieval on Gold.}
\label{tab:rq3}
\footnotesize
\setlength{\tabcolsep}{2.2pt}
\begin{tabular}{@{}lrrrrrrr@{}}
\toprule
 & \multicolumn{5}{c}{1:10} & 1:100 & 1:500 \\
\cmidrule(lr){2-6}\cmidrule(lr){7-7}\cmidrule(l){8-8}
Method & MRR & \begin{tabular}[b]{@{}c@{}}Hits\\@1\end{tabular} & \begin{tabular}[b]{@{}c@{}}Hits\\@5\end{tabular} & \begin{tabular}[b]{@{}c@{}}Article-\\level\\recall\end{tabular} & MRR$^\dagger$ & MRR & MRR \\
\midrule
LODESTONE & \textbf{98\%} & \textbf{96\%} & \textbf{100\%} & \textbf{82\%} & \textbf{95\%} & \textbf{95\%} & \textbf{91\%} \\
\midrule
BM25~\cite{robertson2009bm25} & 89\% & 85\% & 93\% & 74\% & 84\% & 87\% & 82\% \\
\shortstack[l]{Query likelihood\\$\mu=2000$~\cite{zhai2001smoothing}} & 62\% & 48\% & 85\% & 43\% & 56\% & 56\% & 53\% \\
\shortstack[l]{Query likelihood\\$\mu=500$} & 61\% & 48\% & 74\% & 44\% & 55\% & 51\% & 45\% \\
BGE-M3, frozen~\cite{chen2024bgem3} & 28\% & 19\% & 33\% & 19\% & 23\% & 18\% & 15\% \\
Sentence-BERT~\cite{reimers2019sbert} & 39\% & 33\% & 46\% & 26\% & 39\% & 27\% & 21\% \\
POIROT~\cite{milajerdi2019poirot} & 56\% & 52\% & 59\% & 40\% & 57\% & 39\% & 26\% \\
\midrule
DeepHunter~\cite{wei2021deephunter} & 36\% & 21\% & 53\% & 24\% & 33\% & 18\% & 14\% \\
MEGR-APT~\cite{aly2024megrapt} & 1\% & 0\% & 1\% & 0\% & 2\% & 0\% & 0\% \\
NeuroMatch~\cite{ying2020neuromatch} & 1\% & 0\% & 1\% & 0\% & 2\% & 0\% & 0\% \\
ProvG-Searcher~\cite{altinisik2023provgsearcher} & 11\% & 10\% & 12\% & 8\% & 6\% & 5\% & 3\% \\
APT-CGLP~\cite{qiu2026aptcglp} & 1\% & 0\% & 1\% & 0\% & 1\% & 0\% & 0\% \\
\bottomrule
\end{tabular}
\end{table}

Free of localization errors, with human-verified attack edges as queries in the 713-KG pool, LODESTONE ranks a source article first in 96\% of the cases and in the top five in every case (Table~\ref{tab:rq3}). Its MRR and article-level recall are 98\% and 82\%, against 89\% and 74\% for BM25, the strongest retrieval baseline. Lexical retrieval methods (BM25, query likelihood) rank KGs by the terms they share with the query, so they credit the indicator names a PG shares with its source article but cannot relate an operation such as a rename to the behavior an article calls encryption. Dense retrievers (frozen BGE-M3, fine-tuned Sentence-BERT) compress each graph into one vector, so the few relations that match can be diluted by the rest of the graph; they reach 28\% and 39\%. LODESTONE instead compares nodes and edges across the two graphs and rewards pairs that prefer each other among competitors (Section~\ref{sec:locate-reports}). With LODESTONE's located PG as the query (MRR$^\dagger$), LODESTONE and BM25 reach 95\% and 84\%. We further compare mean average precision (MAP), which counts every source article of a case as relevant: LODESTONE reaches 87\% against 79\% for BM25, 55\% and 52\% for query likelihood, 47\% for POIROT, 25\% for frozen BGE-M3, and at most 9\% for MEGR-APT, NeuroMatch, ProvG-Searcher, and APT-CGLP.

With our adaptations (Appendix~\ref{app:adaptations}), POIROT reaches 56\% MRR, DeepHunter 36\%, and MEGR-APT, NeuroMatch, ProvG-Searcher, and APT-CGLP at most 11\%. A PG and an article-level KG describe the same behavior with different entities, relations, and levels of detail, which makes graph-level comparison hard. In our adaptation, DeepHunter's result depends on reducing the PG with the KG's indicators (1\% MRR without this step). These results describe our adaptations rather than the original methods, which were designed for different tasks, inputs, and training budgets.

LODESTONE keeps 95\% and 91\% MRR among 6{,}363 and 31{,}563 candidate KGs, against 87\% and 82\% for BM25, an advantage of about eight to nine points in all three pools.

\subsection{RQ4: Edge-Level Matching}
\label{sec:results-rq4}

\begin{table}[t]
\centering
\caption{PG--KG edge matching on Gold, with the exemption of Section~\ref{sec:metrics}.}
\label{tab:rq4}
\footnotesize
\setlength{\tabcolsep}{4pt}
\begin{tabular}{@{}lrr@{}}
\toprule
Method & \begin{tabular}[b]{@{}c@{}}KG-edge match\\recall $\uparrow$\end{tabular} & \begin{tabular}[b]{@{}c@{}}Linked\\PG units/case $\downarrow$\end{tabular} \\
\midrule
LODESTONE (Qwen3.5-4B) & \textbf{75\%} & 24.8 \\
\quad with additional stepwise review & 72\% & 15.5 \\
\quad without fine-tuning & 52\% & 16.6 \\
\midrule
BERT pair classifier~\cite{devlin2019bert} & 61\% & 47.5 \\
RoBERTa pair classifier~\cite{liu2019roberta} & 51\% & 25.7 \\
DeBERTa pair classifier~\cite{he2023debertav3} & 44\% & 22.4 \\
BGE-M3 mutual neighbors~\cite{chen2024bgem3} & 28\% & 6.7 \\
POIROT~\cite{milajerdi2019poirot} & 0\% & 0.0 \\
Dual-softmax assignment~\cite{sun2021loftr} & 11\% & 2.1 \\
ThreatRaptor~\cite{gao2021threatraptor} & 0\% & 0.0 \\
FGW~\cite{vayer2019fgw} & 41\% & 33.0 \\
\bottomrule
\end{tabular}
\end{table}

\begin{figure}[t]
\centering
\includegraphics[width=\columnwidth]{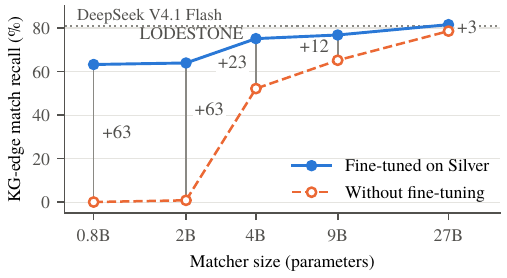}
\caption{KG-edge match recall by matcher size, with and without fine-tuning.}
\label{fig:rq4-size}
\end{figure}

Given the same localized PG units and retrieved KGs, LODESTONE reaches 75\% KG-edge match recall with 24.8 linked PG units per case, against 61\% and 47.5 for BERT, the strongest non-LLM baseline (Table~\ref{tab:rq4}); without the PG-only exemption (Section~\ref{sec:metrics}), they reach 64\% and 46\%. Of the 117 candidate PG units per case, only 6.9 have a verified match. In our adaptations of CTI-query hunting (POIROT, ThreatRaptor), no query is matched in the PG, so neither method links any KG edge. Pair classifiers (BERT, RoBERTa, DeBERTa) score each PG unit--KG edge pair on its own, so a score does not reflect the other KG edges that could explain the same event; BERT links nearly twice as many PG units as LODESTONE yet matches fewer KG edges. LODESTONE instead uses the long context of an LLM to show each PG unit with all candidate KG edges at once and selects a possibly empty set of them (Section~\ref{sec:match-edges}). One-to-one assignment methods (mutual nearest neighbors, dual-softmax assignment) keep only mutual best matches, whereas one PG unit may support several KG edges and one KG edge may need several PG units (Section~\ref{sec:kg-story-relationship}). Soft structural alignment (FGW) reaches 41\% by linking PG units through node similarity and topology, but similar endpoints alone do not show which operation took place (Appendix~\ref{app:adaptations}).

Fine-tuning on Silver raises KG-edge match recall at all five sizes (0.8B, 2B, 4B, 9B, 27B), by about 63, 63, 23, 12, and 3 percentage points (Figure~\ref{fig:rq4-size}). The 0.8B and 2B matchers rise from 0\% and 1\% to 63\% and 64\%: untuned, they mostly pick the first-listed KG edge, so the shuffled options split the votes and almost no link receives four. With the same recipe, Qwen3.8-27B~\cite{qwenteam2026qwen38} reaches 82\%, close to zero-shot DeepSeek V4.1 Flash (81\%), with 22.1 versus 24.6 linked PG units per case. The stepwise reviewer, which also reads adjacent PG edges, cuts linked PG units from 24.8 to 15.5 per case while recall falls from 75\% to 72\%.

\section{Related Work}
\label{sec:related-work}

\noindent\textbf{Provenance-Based Attack Investigation.} Starting from a POI, BackTracker, PrioTracker, and DEPIMPACT trace or prioritize system dependencies~\cite{king2003backtracker,liu2018priotracker,fang2022depimpact}; SLEUTH and HOLMES construct attack scenarios from tag propagation and tactic, technique, and procedure (TTP) patterns~\cite{hossain2017sleuth,milajerdi2019holmes}; ATLAS learns attack patterns~\cite{alsaheel2021atlas}; NODLINK, KAIROS, and ORTHRUS detect anomalies and reconstruct attacks around them~\cite{li2024nodlink,cheng2024kairos,jiang2025orthrus}; and TREC recognizes tactics and techniques from provenance subgraphs~\cite{lv2024trec}. They output a dependency subgraph, a scenario graph, or a technique label; none selects, from a broad CTI knowledge base, the articles describing the observed steps or writes a report citing them. LODESTONE does both, and its training and test cases take security knowledge from CTI through RSI and human curation, respectively.

\noindent\textbf{CTI-Guided Investigation.} EXTRACTOR, AttacKG, and CTINexus turn CTI text into graphs~\cite{satvat2021extractor,li2022attackg,cheng2025ctinexus}. POIROT and ThreatRaptor hunt for a given CTI query in audit data~\cite{milajerdi2019poirot,gao2021threatraptor}; KnowHow matches system events against indicators extracted from CTI and organized by ATT\&CK technique~\cite{meng2026knowhow}; APT-CGLP retrieves CTI for a provenance graph by graph--text matching~\cite{qiu2026aptcglp}; and OCR-APT writes a report from an anomalous subgraph~\cite{aly2025ocrapt}. They hunt for a given query, retrieve articles, label events with ATT\&CK techniques, or write narratives; none provides verified edge-to-edge correspondences between article relations and PG events for training and evaluation, as Gold and Silver do.

\noindent\textbf{Graph Matching and Alignment.} NeuroMatch~\cite{ying2020neuromatch} decides subgraph relations between graphs, the provenance-graph matchers DeepHunter, MEGR-APT, and ProvG-Searcher (Section~\ref{sec:baselines}) compare a PG with query graphs expressed in its system entity and event types, and SuperGlue and LoFTR~\cite{sarlin2020superglue,sun2021loftr} match two sets of visual features. The article-level KGs used here, however, keep high-level entities, such as malware families, and free-form relations. LODESTONE instead fine-tunes a general-purpose LLM on CTI-grounded correspondences, combining security knowledge with the LLM's general abilities to match across vocabularies and graph structures.

\section{Discussion}
\label{sec:discussion}

\noindent\textbf{Synthetic Data.} LODESTONE uses LLM-driven RSI to synthesize part of its training material. Its KGs are extracted from real CTI reports, and all PGs are constructed from logs of software actually executed in VMs. Inspired by synthetic-data research~\cite{yang2025swesmith,qiu2026aptcglp}, we will investigate whether synthetic KGs can improve PG--KG matching.

\noindent\textbf{Skill Evolution.} Recent studies show that iterating external skills can improve LLM systems without weight updates~\cite{zhang2025ace,wang2026valve}. We use Silver's edge-level annotations to train the matching model directly. Skill evolution is a complementary route; future work will compare and combine the two.

\noindent\textbf{End-to-End Reinforcement Learning of Agent Loops.} We considered an agent starting from POIs with only a final report score (0--100\%) as reward. We did not adopt it because terminal feedback gives little guidance about individual decisions across many agent loops. Future work will examine stage rewards and learning from local retries at identified failure points~\cite{shi2026r3l}.

\section{Conclusion}
\label{sec:conclusion}

We presented LODESTONE, a PG--KG graph alignment system that connects CTI expert knowledge to system provenance for evidence-grounded cyber threat investigation. Its CTI-driven data engine pairs reported attack behavior with isolated execution traces, using recursive agent verification and repair to construct training correspondences and human verification to establish the evaluation benchmark. A three-step pipeline recovers compact alert-centered PG subgraphs, narrows the knowledge search through vision-inspired KG retrieval, and applies a fine-tuned LLM to establish partial, many-to-many event-to-relation links. These links support attack summary graphs whose interpretations are traceable to observed events and CTI source sentences.

On 27 Gold cases and a knowledge base of 713 CTI articles, LODESTONE's reports recover 70\% of reference attack facts, achieve an 82\% attack-step description score, and identify the attack in 91\% of cases. Compared pipelines and general-purpose investigation agents reach description scores of at most 57\%. Evidence ablations show that explicit alignment improves report completeness beyond supplying retrieved article text alone. Together, these results demonstrate the value of grounding CTI knowledge in concrete execution evidence to help investigators explain suspicious activity and reconstruct attack steps.

\bibliographystyle{IEEEtran}
\bibliography{references}

\appendices

\section{Adaptations of Compared Methods}
\label{app:adaptations}

Table~\ref{tab:adaptations} lists the adaptations. DEPCOMM's RQ2 results cover the 25 of 27 cases that finish within six hours with its path search capped at 400{,}000 expansions; uncapped, only 17 finish in that time, with the same coverage on them.

\begin{table*}[t]
\centering
\caption{Adaptations of compared methods: reasons and implementation.}
\label{tab:adaptations}
\footnotesize
\begin{tabular}{@{}p{0.17\textwidth}p{0.35\textwidth}p{0.42\textwidth}@{}}
\toprule
Method & Reason for adaptation & Our implementation \\
\midrule
BackTracker, PrioTracker, DEPIMPACT & The original methods consume audit records; LODESTONE supplies a common PG representation. & Run their tracing algorithms on the PG; extend BackTracker's filtering rules to Windows objects. \\
SLEUTH~\cite{hossain2017sleuth} & The paper mentions two additional confidentiality-tag initialization policies without specifying them. & Define initial tags for credential stores, shell histories, and user documents. \\
DEPCOMM & The original output is a summary of communities and information paths; RQ2 requires PG edges. & Map the summary back to PG edges; bound path expansion under our compute budget. \\
ORTHRUS, TREC & RQ2 starts from supplied POIs and evaluates subgraph localization. & Use ORTHRUS's reconstruction stage and TREC's subgraph sampler for the given POIs. \\
ATLAS & It scores attack entities; RQ2 evaluates edges. & Assign each edge the maximum score of its two endpoints. \\
\midrule
POIROT & KGs lack consistent entity types and hand-written alignment patterns; their edge directions do not denote information flow. & Use token-based candidate alignment, omit type constraints, and include both KG-edge directions. In RQ3, use the alignment score for ranking without its acceptance threshold. \\
ThreatRaptor & Its joint queries target its own audit database and a predefined set of operations. & Execute the joint queries over PG units; generate queries for operations supported by its query language. \\
FGW & The solver aligns nodes, and PG and KG edges use different schemas. & Compare node features and undirected topology, then project node alignments onto edge matches. \\
\midrule
MEGR-APT & Training requires graph-edit-distance supervision for the graphs being compared. & Generate graph pairs and topology-based edit-distance labels from benign PG windows, following the original recipe. \\
ProvG-Searcher & PG and KG nodes share no identities, and KG entity types are incomplete. & Infer missing types and approximate the original intersection score with relation-category coverage. \\
APT-CGLP~\cite{qiu2026aptcglp} & We did not obtain the authors' checkpoint or synthesized training pairs; its text input is a CTI report. & Train on graph--text pairs synthesized from benign windows; represent candidate KGs by their edge sentences. \\
NeuroMatch & PG/KG nodes carry text attributes; our training budget is smaller than the original schedule. & Encode nodes with BGE-M3; use a reduced schedule and select the epoch on Silver validation. \\
DeepHunter~\cite{wei2021deephunter} & We did not obtain an official implementation, and the paper leaves some hyperparameters unspecified. & Re-implement the published architecture, document choices for unspecified settings, and reduce PGs using KG indicators. \\
\bottomrule
\end{tabular}
\end{table*}

\section{Safe and Faithful Reproduction}
\label{app:safe-implement}

Each reproduced behavior must pass two gates. It is \emph{safe}: isolated, controlled, reversible, and non-destructive. It is \emph{faithful}: the PG shows the processes, files, registry entries, domains, addresses, and ports that the relation names, so that results on the benchmark carry over to real incidents. A behavior that fails a gate is replaced by a benign surrogate that leaves the same trace (Table~\ref{tab:safety}); a step is withheld, with the reason recorded, only when no such surrogate exists.

\begin{table}[t]
\centering
\footnotesize
\caption{Safe surrogates that preserve the trace.}
\label{tab:safety}
\begin{tabular}{@{}p{0.27\columnwidth}p{0.65\columnwidth}@{}}
\toprule
Behavior & Surrogate \\
\midrule
Destruction, encryption & Acts only on canary (decoy) files and workspaces created by the run; reversible. \\
Credential and data theft & Reads only dummy credentials, cookies, and documents prepared for the run. \\
Exfiltration, command and control & Sends only to a responder that plays the remote server; no route to the Internet. \\
Persistence & Starts from a clean snapshot; clean-up reverts every service, task, registry entry, file, and setting created. \\
Privilege escalation & Not performed. \\
Lateral movement, escape & Only a directed connection to one designated neighbor machine inside the sandbox. \\
Injection, hooking & Cooperative loading or inter-process communication inside a dummy process; the reduced mechanism is recorded. \\
\bottomrule
\end{tabular}
\end{table}

\noindent\textbf{Specification and code.} Before code generation, each KG relation's specification records the original behavior, surrogate, prerequisites, expected telemetry, and clean-up check. The acting process must match the relation's subject: if malware deletes shadow copies, its own process deletes them. An entity in several relations is played by one process and its descendants, keeping its behaviors connected in the PG.

\noindent\textbf{Network targets.} A domain named in a report resolves inside the sandbox to a crafted address in a reserved, non-routable range, where the responder answers (Table~\ref{tab:network-targets}). When a process resolves a domain and, within 60\,s, connects to an address in the answer, its operations on that connection are also recorded as derived edges to the domain; the address edges remain, and ambiguous cases get no domain edge.

\begin{table}[t]
\centering
\footnotesize
\caption{What the PG records for each kind of network target.}
\label{tab:network-targets}
\begin{tabular}{@{}p{0.38\columnwidth}p{0.55\columnwidth}@{}}
\toprule
Target in the report & What the PG records \tabularnewline
\midrule
\raggedright A domain & \raggedright The domain and its sandbox address \tabularnewline
\raggedright A public IP address & \raggedright That IP address \tabularnewline
\raggedright An unnamed server, e.g., ``a control server'' & \raggedright The address of the sandbox responder \tabularnewline
\raggedright Local communication & \raggedright The loopback address \tabularnewline
\bottomrule
\end{tabular}
\end{table}

\noindent\textbf{Background activity.} \label{app:background-activity}
In noisy executions, a persistent worker in a continuously running virtual machine draws an activity category, a real program, one of its tasks, and a configuration (Table~\ref{tab:noise-programs}), re-draws at random intervals, and logs every task it starts. Gold and Silver favor different categories and programs.

\begin{table}[t]
\centering
\scriptsize
\caption{Background activity categories and example programs.}
\label{tab:noise-programs}
\begin{tabular}{@{}p{0.34\columnwidth}p{0.58\columnwidth}@{}}
\toprule
Category & Programs \\
\midrule
Web browsing & w3m, Microsoft Edge, curl \\
E-mail & Mailpit, curl (SMTP), swaks \\
Messaging & curl, netcat, PowerShell (IRC) \\
Source control & Git, Mercurial, Fossil \\
Building & GCC, GNU Make, MSBuild, C\# and Visual Basic compilers \\
Databases & SQLite, DuckDB, Redis, Berkeley DB \\
Documents & Vim, Pandoc, PowerShell \\
Archiving, backup & GNU tar, Info-ZIP, 7-Zip \\
Local web services & nginx, IIS, Python http.server \\
Scheduling & cron, systemd timers, Task Scheduler \\
Monitoring, logs & journalctl, wevtutil, Get-WinEvent \\
Media & VLC, FFmpeg, ffprobe \\
Text search & ripgrep, jq, Select-String \\
Network administration & OpenSSH, ss, Test-NetConnection \\
System inspection & ps, tasklist, whoami \\
File operations & cp, sha256sum, robocopy, xcopy \\
\bottomrule
\end{tabular}
\end{table}

\section{Ethics Considerations}
\label{app:ethics}

\noindent\textbf{Contained reproduction.} We reproduce behaviors from published CTI reports only in our own virtual machines with no route to the Internet, replacing harmful effects with harmless stand-ins that leave the same trace (Appendix~\ref{app:safe-implement}, Table~\ref{tab:safety}). No malware is executed, every run is rolled back, and no system outside our control is contacted.

\noindent\textbf{No new attacks, no human subjects.} All reproduced behaviors are already public; we develop no new attack and find no vulnerability. The work involves no human subjects and no personal data: background activity comes from scripted programs, not from real users.

\section{Use of Generative AI}
\label{app:genai}

\noindent\textbf{In the research.} LLMs extract KGs from CTI articles. These KGs are the fixed input of LODESTONE and every compared method: the reproduction executes the behaviors a KG describes and experts verify its edges' correspondences to the PG, so RQ1--RQ4 measure alignment to the KG as given, not how faithfully the extraction summarizes the article. LLM coding agents (OpenAI Codex and Anthropic Claude Code) write the reproduction plans and code under high-level human guidance; the agents that interact with the sandboxed reproductions run GPT-6.1 Sol in Codex and Claude Opus 4.8 and Claude Sonnet 5.5 in Claude Code. After the reproductions, outside the sandbox, Claude Opus 5.5 revised once the PG-unit-to-KG-edge labels of the Silver records used to fine-tune the matcher, following the matcher's task definition. Humans validate Gold reproductions on their execution evidence and confirm Gold's reference ASGs; agents build Silver cases and their ASGs, which must pass automatic checks. LODESTONE's matcher is the open-weight Qwen3.5-4B fine-tuned with LoRA, and the open-weight GLM-5.3-Flash generates the reports. DeepSeek-V4.1-Flash at temperature 0 is the judge (Section~\ref{sec:metrics}).

\noindent\textbf{Reproducibility and compute.} Because commercial models are not deterministic, all experiments use frozen AI-generated KGs, reproduction code, and labels. We used NVIDIA RTX A6000, RTX 6000 Ada, A100, and GH200 GPUs and Intel Gaudi HL-225D accelerators; for anonymity, the computing platforms and the cloud providers hosting the reproductions will be named in the camera-ready version.

\noindent\textbf{Cost.} Investigating one Gold case takes 6--10 GPU-minutes, almost all in the 4B matcher's eight votes per PG unit (266--557\,s on an A100); replayed prompts show that LODESTONE reads and writes 60k tokens per case, including report generation, against 666k--1.03M for the SWE-agent baselines.

\noindent\textbf{Writing.} Generative AI was used for editorial purposes in this manuscript. The authors gave the editing instructions, inspected all outputs for accuracy and originality, and confirmed that the edited text reflects their intent.

\section{Building a Provenance Graph from Host Events}
\label{app:pg-construction}

We build the PG from events recorded by Windows Performance Recorder (Event Tracing for Windows) or, on Linux, by Sysdig. Nodes represent the processes, files, registry keys and values, domains, and network endpoints named in the trace. Each edge records an operation and its time and follows the data flow: when Word reads \texttt{Invoice.docm} in the Dridex case, the edge points from the file to the process, and a write points from the process to the file. The matcher reads this edge as source \texttt{Invoice.docm}, operation \texttt{FileIo/Read}, and destination \texttt{WINWORD.EXE} (names shortened). Table~\ref{tab:pg-ambiguities} shows how details in a Windows event determine the operation or endpoint. On Linux, refused connections and failed file accesses are left out of the PG.

\section{Linking KG Edges to PG Edges}
\label{app:correspondence}

For each KG relation, we label the PG events that show the action the report states. Suppose an installer opens \texttt{helper.exe}, writes it to disk, and then starts it. For \emph{the installer dropped \texttt{helper.exe}}, we label the writes; if it wrote the file in three chunks, all three writes count. For \emph{the installer executed \texttt{helper.exe}}, we label the process start. Opening the file alone establishes neither claim.

\begin{table}[t]
\centering
\footnotesize
\caption{How a Windows trace event becomes a PG edge.}
\label{tab:pg-ambiguities}
\begin{tabular}{@{}p{0.31\columnwidth}p{0.63\columnwidth}@{}}
\toprule
\raggedright Windows event & \raggedright What the graph records \tabularnewline
\midrule
\raggedright File write & \raggedright An ordinary write is labeled \emph{Write}; a flush of cached bytes to disk carries the paging flag and is labeled \emph{WritePaging}. \tabularnewline
\raggedright File create request & \raggedright A request only to open an existing file is labeled \emph{Open}; other modes keep \emph{Create}, which alone does not show a new file. \tabularnewline
\raggedright Directory listing & \raggedright The endpoint is the searched directory; a pattern such as \texttt{*.txt} is stored on the edge. \tabularnewline
\raggedright Process termination & \raggedright The edge points from the caller to the target, whose exit is recorded separately. \tabularnewline
\raggedright TCP connection accepted & \raggedright The edge points from the server's listening address and port to the accepting process. \tabularnewline
\bottomrule
\end{tabular}
\end{table}

The agent labels events with Table~\ref{tab:match-rules} as a reference and can also implement other reported behaviors in code and label their PG edges. The selected edges need not form a connected path. These direct matches are \emph{normal} edges; preparatory or incidental operations, such as the queries around a registry write, are labeled separately as \emph{weak} edges. In Gold, curators confirm both kinds, and weak edges are kept for future research. This paper uses only normal edges: they are the attack edges of RQ2, and RQ4 scores them as PG units; the three writes above form one PG unit if their source--operation--target text is identical. Table~\ref{tab:exceptions} covers repeated actions and the object named in the report.

\begin{table}[t]
\centering
\footnotesize
\caption{Recorded events that support each claim, with Windows operation names shortened.}
\label{tab:match-rules}
\begin{tabular}{@{}p{0.41\columnwidth}p{0.53\columnwidth}@{}}
\toprule
\raggedright Action in the KG & \raggedright Direct evidence in the PG \tabularnewline
\midrule
\raggedright Drops or creates a file & \raggedright \emph{Write} to the named file \tabularnewline
\raggedright Encrypts a file & \raggedright \emph{Write} to the encrypted output \tabularnewline
\raggedright Reads a target file (e.g., to collect or steal it) & \raggedright \emph{Read} from the named file \tabularnewline
\raggedright Copies or decodes a file & \raggedright \emph{Read} from the input and \emph{Write} to the output \tabularnewline
\raggedright Deletes a file & \raggedright \emph{Delete} and \emph{DeletePath} for the file \tabularnewline
\raggedright Empties a file & \raggedright A \emph{Write} that leaves the file empty \tabularnewline
\raggedright Renames a file or changes its extension & \raggedright \emph{Rename} for the old name and \emph{RenamePath} for the new name \tabularnewline
\raggedright Searches a directory & \raggedright \emph{DirEnum} on the searched directory \tabularnewline
\raggedright Creates a file in a named directory & \raggedright \emph{Create} on the directory named in the claim (see Table~\ref{tab:pg-ambiguities}) \tabularnewline
\midrule
\raggedright Runs a program (executes, deploys, triggers) & \raggedright \emph{Process/Start} from the parent to the new process \tabularnewline
\raggedright Loads a module & \raggedright \emph{Image/Load} for that module \tabularnewline
\raggedright Uses a Windows API & \raggedright \emph{Image/Load} for an API-specific module; individual calls are not visible. \tabularnewline
\raggedright Terminates a process & \raggedright \emph{Terminate} from caller to target and \emph{End} on the target \tabularnewline
\raggedright Persists via a registry value & \raggedright \emph{SetValue} on the named registry value \tabularnewline
\midrule
\raggedright Downloads a file & \raggedright \emph{Recv} of file data from the source and \emph{Write} to the destination file \tabularnewline
\raggedright Communicates with a peer & \raggedright Every \emph{Send} to or \emph{Recv} from that peer \tabularnewline
\raggedright Sends data to a peer & \raggedright Only \emph{Send} to that peer \tabularnewline
\raggedright Uses a network port & \raggedright \emph{Connect} or \emph{Reconnect} to that port \tabularnewline
\bottomrule
\end{tabular}
\end{table}

\begin{table}[t]
\centering
\footnotesize
\caption{How to label repeated actions and the named object.}
\label{tab:exceptions}
\begin{tabular}{@{}p{0.34\columnwidth}p{0.60\columnwidth}@{}}
\toprule
\raggedright What the annotator sees & \raggedright Which events to label \tabularnewline
\midrule
\raggedright A report names a group of files & \raggedright For \emph{writes ransom notes}, label the writes to each note. A write can also support a relation naming that individual file. \tabularnewline
\raggedright One registry value changes three times & \raggedright In Dridex, the first write to \texttt{OfficeUpdate} supports \emph{adds}, the deletion \emph{removes}, and the second write \emph{adds again}. \tabularnewline
\raggedright The report names the parent registry key & \raggedright In the same example, label the change to the value \texttt{OfficeUpdate}, not its key \texttt{Run}; opening the key does not show the change. \tabularnewline
\raggedright A file write appears as a paging write & \raggedright Include the flush attributed to the same process and file. With memory-mapped I/O this may be the only write. Exclude a flush attributed to another process. \tabularnewline
\raggedright The report names a domain & \raggedright Use the edge to the domain when DNS ties the connection to it (Table~\ref{tab:network-targets}); a DNS lookup alone shows only the lookup. \tabularnewline
\raggedright Names differ between machines & \raggedright The local user, host or folder may differ. A domain, address or malicious filename specified as an indicator must match. \tabularnewline
\bottomrule
\end{tabular}
\end{table}

\noindent\textbf{When no event shows the action.} A read of the document \texttt{Invoice.docm} cannot establish the claim that it contained a macro, and for code running inside another process, a nearby process start is at most a weak edge. Before accepting a gap, the agent rules out the four sources of missing PG evidence in Table~\ref{tab:missing-pg-evidence}. If the gap remains, the KG relation is assigned the first applicable class in Table~\ref{tab:not-implementable}.

\begin{table}[t]
\centering
\scriptsize
\caption{Classes of KG relations without a safe, observable PG counterpart.}
\label{tab:not-implementable}
\begin{tabular}{@{}p{0.30\columnwidth}p{0.62\columnwidth}@{}}
\toprule
Class & Criterion \\
\midrule
A. Withheld by the safety gate & A harmful action, e.g., contacting a public control server or reading another process's memory. \\
B. An endpoint has no PG object & Content inside a file or registry value (the write itself is evidenced by another relation); free-standing data; an abstract entity, such as a malware family or a party behind an anonymity network; an attribute of a node, such as \emph{is\_a}; a delivery channel, such as e-mail. \\
C. No operation between PG entities & A change inside process memory with no process start before or after it; a display or a user interaction. \\
\bottomrule
\end{tabular}
\end{table}

\section{Judge Agreement}
\label{app:judge-agreement}

\begin{table}[t]
\centering
\scriptsize
\setlength{\tabcolsep}{5pt}
\caption{Pairwise agreement (\%) on descriptions (below the diagonal) and facts (above).}
\label{tab:judge-agreement}
\begin{tabular}{@{}lcccccc@{}}
\toprule
 & Judge (DS) & GLM & Sonnet & A & B & C \\
\midrule
Judge (DS) & -- & 94 & 94 & 98 & 96 & 88 \\
GLM & 92 & -- & 100 & 96 & 98 & 86 \\
Sonnet & 98 & 94 & -- & 96 & 98 & 86 \\
A & 90 & 86 & 88 & -- & 98 & 86 \\
B & 88 & 82 & 86 & 92 & -- & 88 \\
C & 88 & 84 & 86 & 86 & 84 & -- \\
\bottomrule
\end{tabular}
\end{table}

\noindent\textbf{Several facts per step.} In the expert review, some facts were labeled unmatched only because the report generator stated several facts in one step, against its instruction to state one fact per step; this is an error of the generating LLM. We therefore also ran the judge with a step allowed to match several facts. On the reviewed facts, this judge agrees with GLM-5.3-Flash, Claude Sonnet 5.5, and authors A, B, and C on 96\%, 92\%, 90\%, 90\%, and 82\% of labels. Under it, LODESTONE's fact recall is 74\% (70\% in Table~\ref{tab:rq1}), still above the stepwise-review variant (67\%), the evidence ablations (67\% and 66\%), and every general-purpose agent (at most 54\%).

\end{document}